\documentclass[twocolumn]{aastex61}
\pdfoutput=1 
\usepackage{amsmath,amstext}
\usepackage[T1]{fontenc}
\usepackage{apjfonts} 
\usepackage[figure,figure*]{hypcap}
\usepackage{tabularx}
\usepackage[colorinlistoftodos]{todonotes}
\usepackage{graphicx}
\usepackage{CJKutf8}

\usepackage[T1]{fontenc}
\usepackage[utf8]{inputenc}

\usepackage{listings}

\DeclareFixedFont{\ttb}{T1}{txtt}{bx}{n}{10} 
\DeclareFixedFont{\ttm}{T1}{txtt}{m}{n}{10}  

\usepackage{color}
\definecolor{deepblue}{rgb}{0,0,0.0}
\definecolor{deepred}{rgb}{0.6,0,0}
\definecolor{deepgreen}{rgb}{0,0.5,0}

\usepackage{listings}

\newcommand\pythonstyle{\lstset{
language=Python,
basicstyle=\ttm,
otherkeywords={self},             
keywordstyle=\ttm\color{deepblue},
emph={MyClass,__init__},          
emphstyle=\ttb\color{deepred},    
frame=tb,                         
showstringspaces=false            %
}}

\newcommand\py[1]{{\pythonstyle\lstinline!#1!}}

\usepackage{natbib}
\usepackage{har2nat}

\shorttitle{First Principle Simulator}
\shortauthors{Dodkins et al.}

\begin{document}


\title{First Principle Simulator of a Stochastically Varying Image Plane for Photon-Counting High Contrast Applications}

\author{Rupert H. Dodkins}
\affiliation{Department of Physics, University of California, Santa Barbara, California 93106, USA}
\author{Kristina K. Davis \footnote{NSF Astronomy and Astrophysics Postdoctoral Fellow}}
\affiliation{Department of Physics, University of California Santa Barbara, Santa Barbara, California 93106, USA}
\author{Briley Lewis}
\affiliation{Division of Astronomy and Astrophysics, University of California, Los Angeles, Los Angeles, California 90095, USA}
\author{Sumedh Mahashabde}
\affiliation{Department of Microtechnology and Nanoscience, Chalmers University of Technology, Gothenburg 412 96, Sweden}
\author{Benjamin A. Mazin}
\affiliation{Department of Physics, University of California Santa Barbara, Santa Barbara, California 93106, USA}
\author{Isabel A. Lipartito}
\affiliation{Department of Physics, University of California Santa Barbara, Santa Barbara, California 93106, USA}
\author{Neelay Fruitwala}
\affiliation{Department of Physics, University of California Santa Barbara, Santa Barbara, California 93106, USA}
\author{Kieran O'Brien}
\affiliation{Department of Physics, Durham University, South Road, Durham, DH1 3LE, UK}
\author{Niranjan Thatte}
\affiliation{Department of Astrophysics, Denys Wilkinson Building, Keble Road, Oxford, OX1 3RH, UK}

\begin{abstract}

Optical and near-infrared Microwave Kinetic Inductance Detectors, or MKIDs are low-temperature detectors with inherent spectral resolution that are able to instantly register individual photons with potentially no false counts or readout noise. These properties make MKIDs transformative for exoplanet direct imaging by enabling photon-statistics-based planet-discrimination techniques as well as performing conventional noise-subtraction techniques on shorter timescales. These detectors are in the process of rapid development, and as such, the full extent of their performance enhancing potential has not yet be quantified.  

MKID Exoplanet Direct Imaging Simulator, or MEDIS, is a general-purpose end-to-end numerical simulator for high-contrast observations with MKIDs. The simulator exploits current optical propagation libraries and augments them with a new MKIDs simulation module to provide a pragmatic model of many of the degradation effects present during the detection process. We use MEDIS to demonstrate how changes in various MKID properties affect the contrast-separation performance when conventional differential imaging techniques are applied to low-flux, short duration observations.

We show that to improve performance at close separations will require increasing the maximum count rate or pixel sampling when there is high residual flux after the coronagraph. We predict that taking pixel yield from the value achieved by current instruments of 80\% and increasing it to 100\% would result in an improvement in contrast of a factor of $\sim$ 4 at 3$\lambda/D$ and $\sim$ 8 at 6$\lambda/D$. Achieving better contrast performance in this low flux regime would then require exploiting the information encoded in the photon arrival time statistics.

 
\end{abstract}


\section{Introduction}
\label{sec:intro}



High contrast imagers use extreme adaptive optics (AO) to provide a high-fidelity correction of the incoming wavefronts, resulting in Strehl ratios of more than 80\% at 1 $\mu m$ \cite{sauvage2014wave, sahoo2018subaru}. The residual optical aberrations produce a quasistatic stochastic speckle noise pattern in the image plane that currently limits the minimum detectable mass of exoplanets to several Jupiter masses at 100 mas with 8 m class telescopes. Discovering companions with lower masses necessitates better elimination of these speckles or better identification of the companions in post processing. 

Microwave Kinetic Inductance Detectors (MKIDs) are a highly sensitive cryogenic photodetector \cite{day2003broadband}. The effective band-gap of the superconducting sensing element is many thousands of times smaller than that of a conventional semiconductor-based detector. This enables the energy of each photon to be determined, in real-time, by the magnitude of the response without the use of complex dispersive optics. (More details on the operating principle of MKIDs can be found in section \ref{subsec:detector}.) 

By capturing the speckle pattern with minimal evolution and detector noise, MKID-based high contrast imaging (HCI) subsystems have the capability to provide superior speckle subtraction. These properties will provide superior performance when applying high dispersion coronography \cite{OBrien_KIDSpec_2013, mazin2019optical} as well as enable new techniques of planet discrimination based on the photon arrival time statistics \cite{walter2019stochastic} for very high contrast performance. 



The MKID Exoplanet Camera (MEC; \citet{walter2018mec}) and the Planetary Imaging Concept Testbed Using a Recoverable Experiment – Coronagraph, (PICTURE-C; \citet{mendillo2018picture}) promise to rigorously demonstrate these transformative benefits. However, it may be a number of years before these instruments are fully realized at their full potential. Detector development, for example, is an ongoing iterative process as fabrication recipes are tuned and new materials investigated \citep{coiffard2020characterization}. To compliment these systems, we have developed a general purpose fully end-to-end numerical simulator for high contrast observations with MKID-based HCI systems -- the MKID Exoplanet Direct Imaging Simulator (MEDIS)\footnote{https://github.com/MazinLab/MEDIS}. With MEDIS we can analyse MKID-based systems in a controlled environment under a variety of conditions set by the parameters. We can then investigate how changes in different MKID parameters impact the performance of the current and near-future system to help guide their development. 

This article is laid out as follows: Section \ref{sec:method} details the implementation of the MEDIS package. Section \ref{sec:results} demonstrates the impact of changing various MKID parameters on the contrast performance after applying conventional post-processing techniques to a short observation. Concluding remarks are found in Section \ref{sec:conclusions}. Some of the main control parameters used for this article's investigation are described in the appendix.

\begin{figure}
  \centering
  \hspace{-1cm}
   \includegraphics[width=\columnwidth]{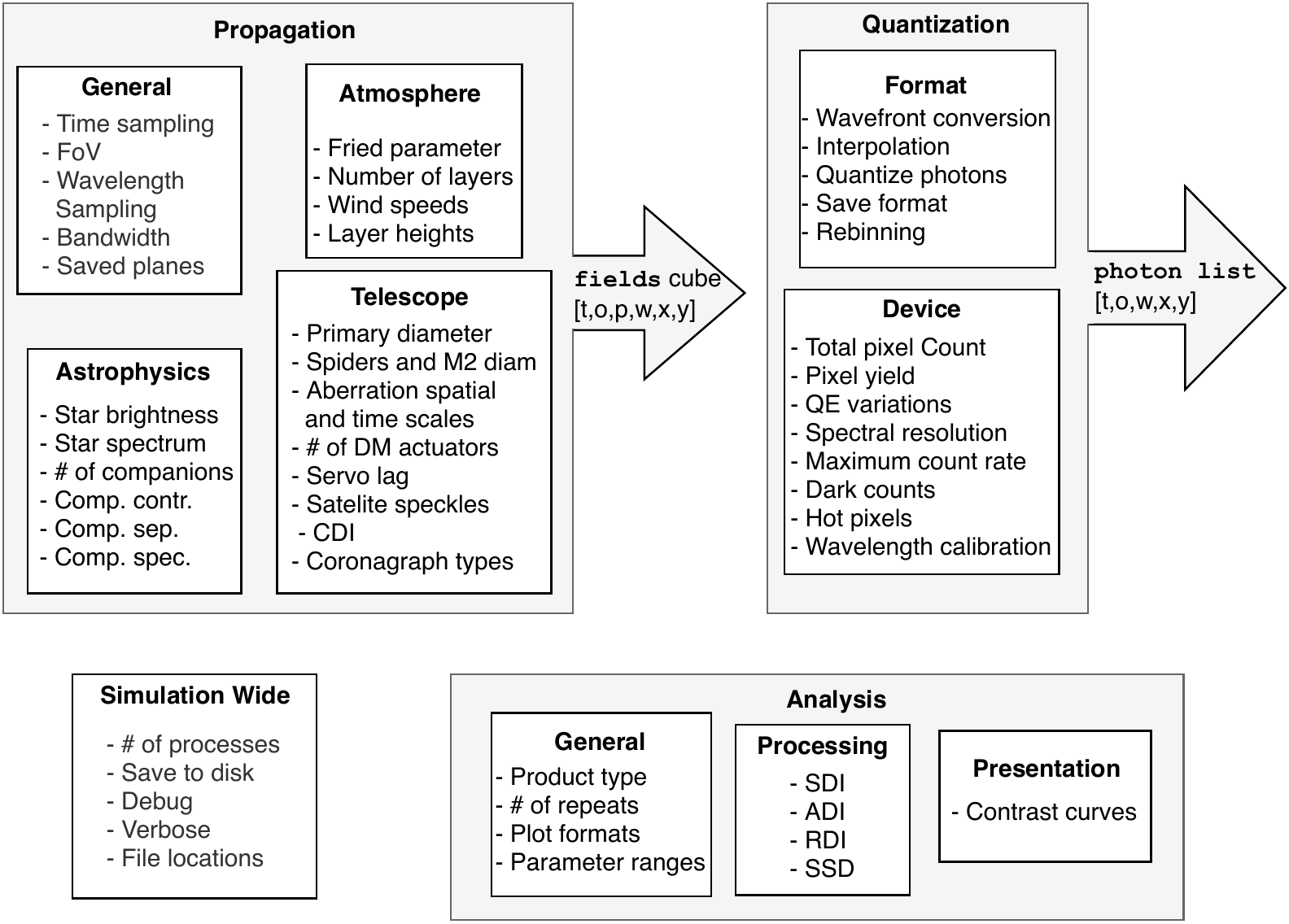}
\hspace{-1cm} 

  \caption{An overview of MEDIS. The arrow shows the direction of information through the pipeline and the format of the information. The data product of Propagation is a six-dimensional tensor of dimensions (arrival time $t$, astronomical object index $o$, optical plane $p$, wavelength $w$, and two spatial dimensions $xy$). The data product of Quantization is an unordered set of photons with attributes (arrival time, wavelength, and two spatial dimensions). Analysis can either produce reduced images or contrast-separation curves. Some of the key parameters for different modules are listed in the white boxes.}
  \label{MEDIS_struct}
\end{figure}

\section{Implementation}
\label{sec:method}

\subsection{Overview}



Figure \ref{MEDIS_struct} is a schematic of the MEDIS pipeline along with some of the relevant parameters at each stage -- some of these main parameters are explained in the appendix. MEDIS first exploits two general-purpose end-to-end numerical simulation packages, PROPER \cite{krist2007proper} and HCIpy \cite{Sebastiaan2018}, to model the initial propagation of a collection of complex electric fields through an optical system up to the detector -- called the \py{Propagation} step. Each field in the collection represents a different astronomical object, wavelength sample, and timestep. At a minimum the fields at the detector plane are stored, but other planes in the optical train can be specified also. Therefore, the data product of \py{Propagation}, called \py{fields}, is a six-dimensional tensor of dimensions: arrival time, astronomical object index, optical plane index, wavelength, and two spatial dimensions. MEDIS can compute the propagation of each wavelength and object field in parallel. It can also compute the timesteps in parallel if the AO configuration allows for it. 

In the \py{Quantization} step, \py{fields} is converted into quantized-photon data as observed by an MKID instrument after detection artifacts have been introduced. The data product of this step, called a \py{photon_list}, is a two-dimensional table with each row representing a new photon and each column representing: arrival time, wavelength, sensor row number, and sensor column number.  These \py{fields} or \py{photon_lists} can then be processed to create images and contrast-separation curves using modules in MEDIS (in future releases these data analysis algorithms will be integrated into MKIDAnalysis pipeline for HCI MKID observation data \cite{mkidanalysis}).

The default configuration parameters that control these steps are contained in a single file called \py{params.py}, and divided into different classes such as `Simulation-wide', `input/output', `Propagation', `telescope', `astrophysics', `atmosphere', `MKID device'. The user then modifies any of these parameters before calling \py{Propagation} or \py{Quantization} to dictate the observation. Instances of \py{Propagation} or \py{Quantization} will then be created and stored along with the parameters used to generate them (what is stored in these realizations is explained in sections \ref{sec:telescope} and \ref{subsec:detector}). MEDIS will automatically use those instances if the parameters match the ones specified by the user on the next call. Similarly, \py{Quantization} will use the stored \py{fields} by default if the relevant parameters match.

\subsection{Propagation}

\label{sec:telescope}
\begin{figure}
  \centering
  \hspace{-0.5cm}
\includegraphics[width=0.46\textwidth]{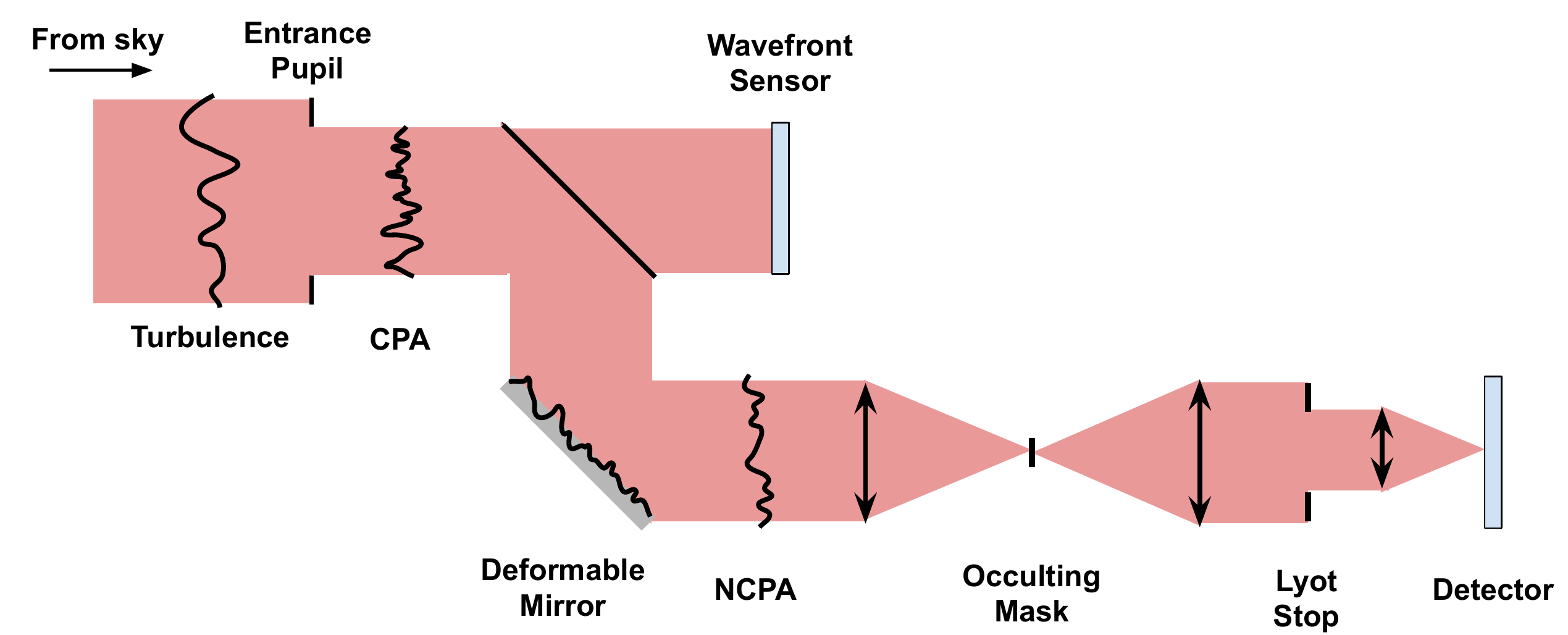}
\includegraphics[width=0.46\textwidth]{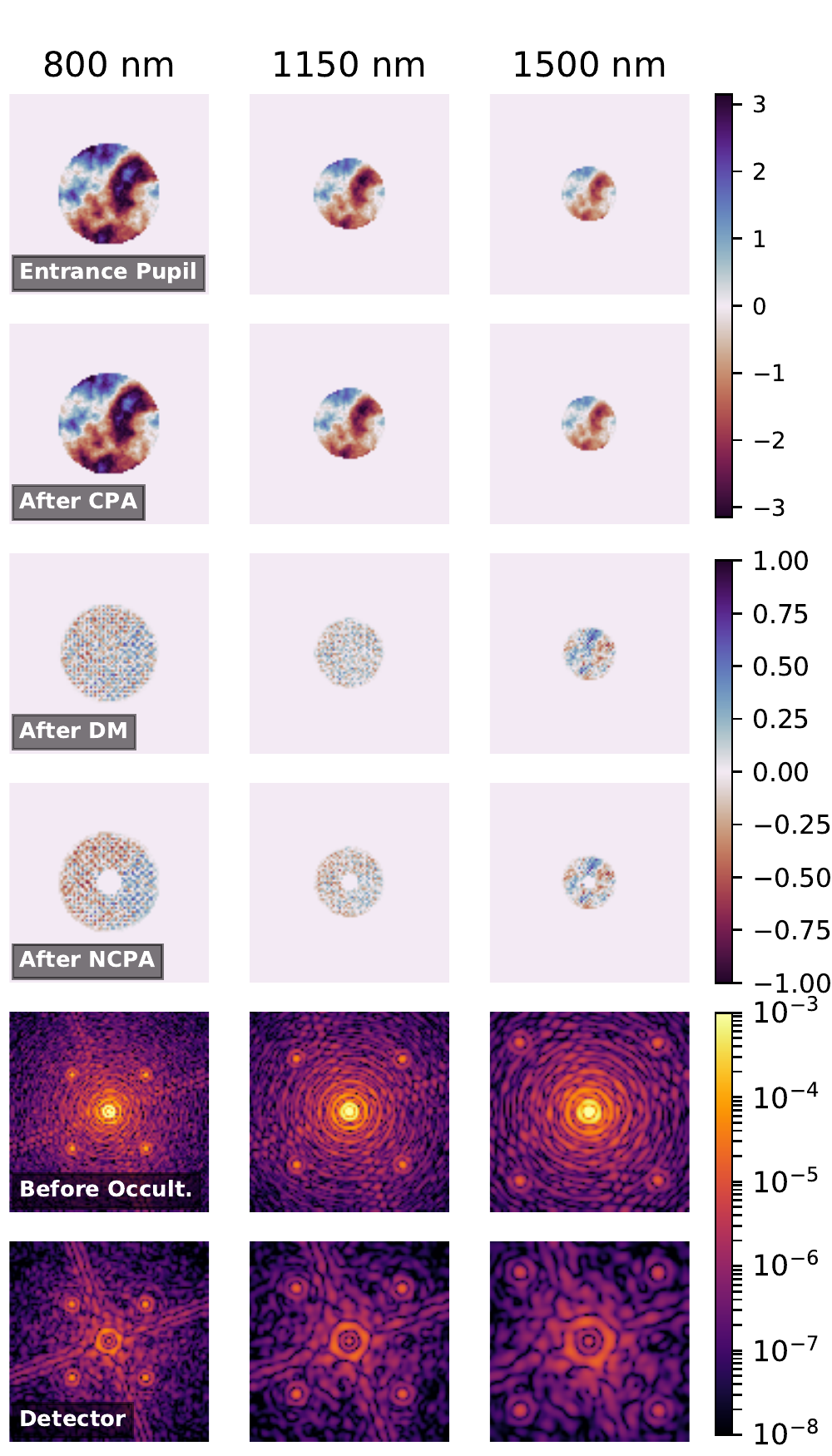}
  \caption{Phase and intensity maps of the primary star electric field at different points in the telescope train obtained from the \py{fields} 6D tensor. The first four rows show phase and the final two rows are intensity.}
  \label{maps}
\end{figure}

PROPER \py{wavefront} objects contains a complex electric field array at a single wavelength. PROPER predicts the diffraction effects on this field in both the near-field and far-field regimes as they propagate through the system. The ability to accurately simulate in the near-field is crucial because high contrast imagers can contain optics outside of the pupil plane. This introduces significant chromatic aberrations into the speckles and makes post processing techniques such as spectral differential imaging (SDI) less effective \citep{marois2006exoplanet}. 

A matrix of PROPER \py{wavefront}s is first generated with dimensions of the number of astronomical objects and wavelength samples. Then, for each timestep, the \py{wavefront}s propagate through a sequence of optical elements determined by a PROPER prescription script (the user can either use one of the examples or they can create their own). Figure \ref{maps} shows the optical train for the `general' prescription, which contains the minimum functionality for a HCI system. Below the prescription schematic are the phase and amplitude maps of the \py{wavefront}s for the primary star at several wavelengths at several planes in the optical train. For the investigations here, all phase aberrations are added in the pupil plane and propagation happens between both pupil and focal planes only.

First, the relative amplitude of the initial wavefronts at different wavelengths for an object is set by the temperature of the object according to the Planck law or loaded reference spectra. The relative amplitude of the wavefronts for different objects is set by the requested contrast ratio (at the shortest wavelength) and spectral type of the objects. For companion objects (not shown in Figure \ref{maps}) a tilt is applied to offset the PSF in the focal plane. 

The \py{wavefront}s pass through relatively quickly evolving atmospheric aberrations before being masked by the entrance pupil, which can be seen in the Entrance Pupil row of Figure \ref{maps}. The atmospheric turbulence is assumed to be localized to a number of thin layers of infinite extent that move independently, and the turbulence within those layers remains frozen for the duration of the observation \cite{taylor1938spectrum}. The user chooses the parameters that generate Kolmogorov power spectrum density (PSD) distributions \cite{kolmogorov1941dissipation} to generate the layers from. For a given time step, the electric field is geometrically propagated through each of the layers and the resulting phase map for each wavelength at the telescope entrance pupil is stored as a FITS file.

The \py{wavefront}s meet another set of aberrations identified here as common path aberrations (CPA), which have a different characteristic spatial frequency, amplitude, and are slowly evolving or static (here the relative amplitudes between the atmospheric and common-path aberrations are such that the effects of CPAs are not that noticeable to the eye). All of the aberration maps are generated before the generation and propagation of the \py{wavefront}s, the first of which are the dynamic aberration maps generated by {\it Atmosphere}. The quasistatic aberration maps are also generated from randomly sampled PSD distributions according to the distribution PSD$_\textrm{2D} = a/(1+(\frac{k}{b})^c)$ \cite{church1991optimal} and can be made to evolve by sampling the phase from a correlated Gaussian distribution.






The \py{wavefront}s then pass through the AO system that senses and corrects most the phase error and adds satelite speckles using the deformable mirror (DM). The WFS takes the phase map of the star at the shortest wavelength and unwraps it. This map is then interpolated onto a grid with the dimensions equal to the number of DM actuators to determine the desired actuator height array. The DM map with the same spatial sampling as the wavefront is then generated accounting for the influence function of the Xinetics Photonex DM \cite{ealey2004high}. A Gaussian uncertainty can then be applied to each actuator in this array to simulate piston error. This DM map is then applied to the 


For a better correction, the AO module can store the DM values, measure the new residual wavefront error and update the DM values accordingly to converge on the minimum wavefront error (at the expense of serializing the simulation processes and dramatically increasing the simulation time).

To introduce servo lag error, a time datacube of WFS measurements is created. For each time-step, the WFS measurement is placed in the datacube at the index corresponding to the time delay. For a time delay $\tau = 10$ ms and the time-step $t =$ 1 ms, this would be the tenth index. The phase map at the first element is read by the DM. All the phase maps then shift one element in the time axis. For $t<\tau$, there has been insufficient time for the phase maps to propagate to first index and the DM receives an array of zeros. At $t\ge\tau$, the DM receives phase maps from time $t - \tau$. To introduce bandwidth error, the phase maps from several elements of the time datacube are mean-averaged before being sent to the DM.

Slowly evolving non-common path aberrations (NCPA) are then applied to the \py{wavefront}s as well as the obscuration of the secondary mirror and the spider legs. Applying the obscurations here instead of the entrance pupil allows for better phase-unwrapping measurements by the wavefront-sensor. 

The final step is a simple Lyot coronagraph that is applied to the \py{wavefront}s before the detector plane. There are a number of occulters that can be selected. 


The pupil and focal planes must have a sufficient number of spatial grid points to sufficiently sample relevant features in each plane. For example two samples are required across the coherence length and the spider arms in the pupil plane. In the focal plane there must be at least one grid pixel for every MKID pixel after the image has been interpolated onto that FoV.  

Increasing the spatial sampling decreases the angular sampling and vice versa. For example, in order to Nyquist sample the focal plane the pupil entrance diameter must be at least half of the grid width at the shortest wavelength. This beam ratio decreases proportionally with increasing wavelength to retain uniform angular sampling in the focal plane between wavelengths. Therefore the grid size must be made ever larger to accommodate for the pupil plane scales at these wavelengths. We found that a grid size of 512 and a FoV of 2.6" is sufficient for a 300x300 pixel MKID array with 10 mas/pixel platescale array (greater than a factor of four above the collecting area of the current MEC devices).

If there are too few grid points across the beam, then the outermost wings of the PSF will cause discontinuities at the boundary, effectively scattering signal into higher spatial frequencies. For the ideal Airy pattern, this manifests as a checkerboard pattern in the focal plane image at large radii.  

The smallest feature for wavelength and time are typically determined by the presence of emission lines (set by the astrophysics parameters) and the fast varying aberration screens (set by the atmosphere parameters).

\subsection{Quantization}
\label{subsec:detector}

\begin{figure*}
  \centering
  \hspace{-1cm}
   \includegraphics[width=\textwidth]{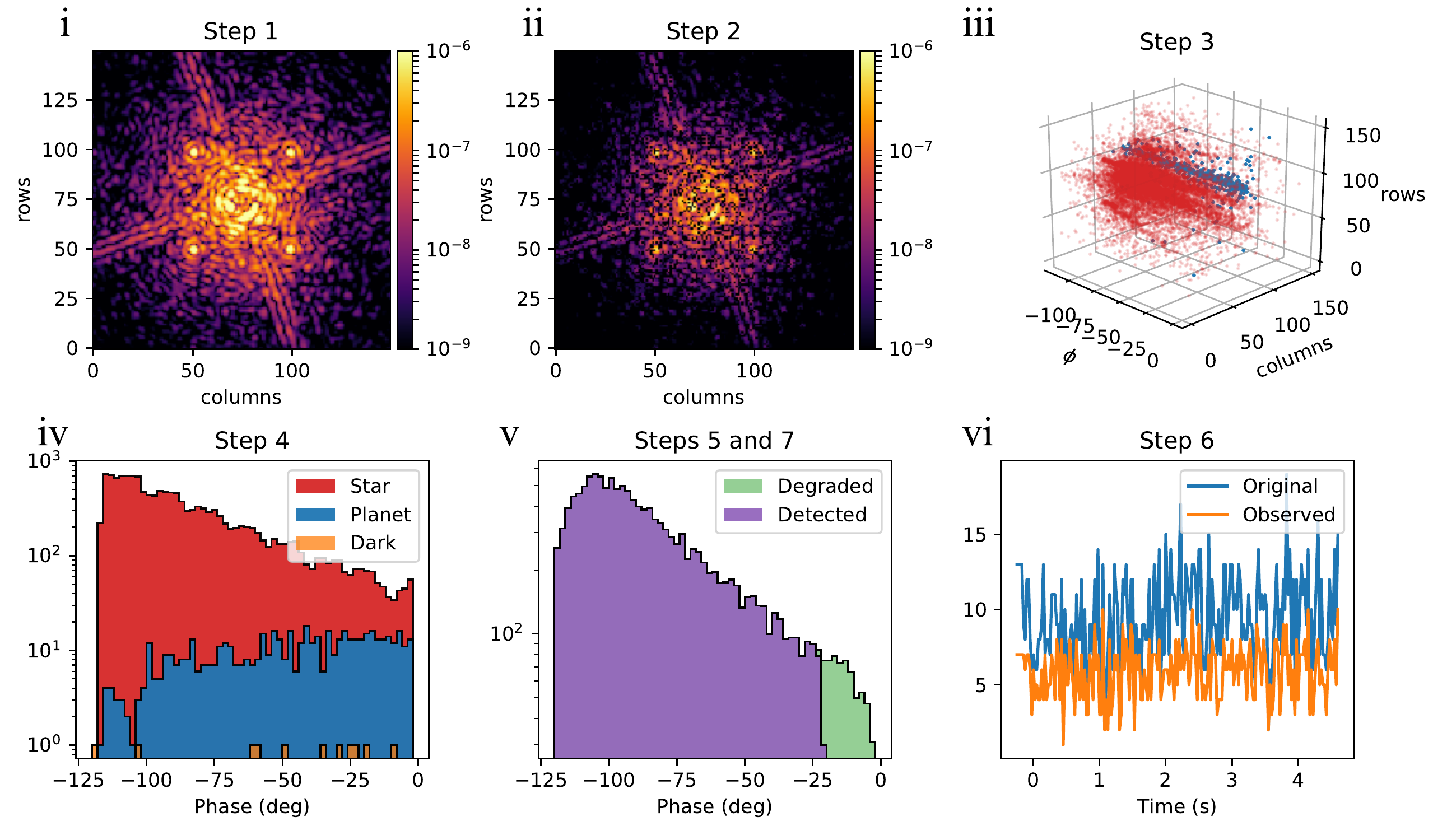}
\hspace{-1cm}
  \caption{Demonstration of the process of creating MKID observations on a star and high mass companion. Panel (i) show the intensity of the primary star electric field at the shortest wavelength at a single timestep. Panel and (ii) shows the result of step 2, which gives the dead pixels their zero intensity. Panel (iii) shows the location of resultant photons, in row column and phase height space, and sampled from the intensity datacube for the star (red) and planet (blue). Panel (iv) is the same data but summed over all spatial dimensions and with the addition of dark photons. In panel (v) all photons are grouped together and degraded according to their pixel's responsivity and $R$ (green). Also in panel (v) are the photons with sufficient phase to be detected (purple). Panel (vi) shows the light curve for an aperture centered on the companion object after many time samples before and after the affect of pixel dead time have been accounted for.}
  \label{idealvsMKID}
\end{figure*}

MKID pixels are microwave resonator circuits etched out of a superconducting material on a substrate and driven by a microwave tone at the resonant frequency. When a photon hits the pixel it breaks many thousands of Cooper pairs through a cascade of phonons, increasing the inductance and decreasing the pixel's resonant frequency, before returning to the equilibrium state. This produces a characteristic pulse shape in that pixel's phase height time stream of the drive tone. The more precisely that pulse height is known, the higher the spectral resolution, $R$. It is calculated from a series of monochromatic photon pulses via $R \approx \phi/\Delta\phi$, where $\phi$ is the mean phase height and $\Delta\phi$ is the FWHM of the Gaussian distribution.

Each pixel has a background phase height noise from two-level system states \citep{gao2008experimental}, and amplifier noise \citep{zobrist2019wide}. Additionally, the phase height responsivity, $r=\lambda/\phi$, varies from pixel to pixel depending on their quality factor and consequently their bias power \cite{dodkins2018mkid}. Only when the phase height signal to noise ratio (S/N) exceeds a defined threshold is the photon data stored to disk by the readout electronics. Longer wavelength photons will have a smaller signal-to-noise ratio and the spectral resolution will decrease towards this regime. 

Currently, MKID devices suffer from dead pixels. Uncertainties in the fabrication cause pixels to be indistinguishable to the readout resulting in a loss of both pixels. Other pixels have dark counts, which typically have an average flux of one per pixel per hour. There is also a minimum time between which two photons can be detected by a the readout. This `dead time' is required because the pixel takes some finite amount of time to reset to equilibrium after detecting the first photon, and subsequent photons impacting the pixel during that time would have an elevated baseline falsely elevating their measured energy. 

These effects have consequences on the performance of MKID instruments for HCI and are therefore introduced in the \py{Quantization} module. First, an instance of \py{Quantization} is created according to the specified parameters. This includes assigning a QE, phase height responsivity, and $R$ to each pixel. These parameters are assumed to have a Gaussian distribution with user defined mean and standard deviation. MEDIS has sensible defaults for these values shown in Table \ref{tab:mkids}, however more accurate values from device measurements should be used for absolute tests of MKID performance. Spectral resolution also has a decreasing dependence with wavelength. The background phase height for each pixel is calculated via it's FWHM at zero phase height $\sigma_{\phi} \approx \Delta\phi(\phi=0)/2.35$. The locations of the dead pixels and hot pixels are also assigned at this time point. Dead pixels are created by setting their responsivity to zero.

This device realisation is then combined with the \py{fields} measurement to create a \py{photon_list}. For each timestep the fields at the detector plane are extracted and converted to intensity thereby generating a spectral datacube for each object. The process of generating a list of photons from these spectral cubes is illustrated in Figure \ref{idealvsMKID} and the steps summarized below:


\begin{enumerate}
\item Select the region of interest that matches the detector FoV and interpolate onto the detector pixel sampling
\item Multiply the datacube by a detector QE map 
\item Draw samples from each object's 3D distribution
\item Add dark photons to the list
\item Multiply the wavelength of each photon by each pixel's responsivity and add a Gaussian random uncertainty bias to the phase heights
\item Remove any photons incident during the dead time of each pixel
\item Remove any photons that have insufficient phase height compared to the pixel noise floor
\end{enumerate}

In Figure \ref{idealvsMKID} the resampling of the spatial dimensions (step 1) and the scaling of the intensity (step 2) are shown for a single wavelength and time. In panel (ii) the device QE map decreases the mean intensity of the image and sets some of the pixels to zero intensity. Panel (iii) then shows the quantization in three-dimensions in the form of a point cloud map. The radial scaling of the PSF towards smaller phase height (longer wavelengths) is observable in this perspective. The fact that the contrast difference between the two objects becomes more favourable towards smaller phase is apparent in both panel (iii) and (iv). Panel (iv) also shows the relatively small contribution of dark counts in MKID devices. Furthermor, the quantization error in the star spectrum of choosing only 8 wavelength samples visible as the discrete jumps in number of counts. Then in panel (v) the degradation of the combined spectra from the responsivity and $R$ variations is shown in the degraded spectra. The Detected curve in the same panel is the result of applying step 7 above. The cut-off in the spectrum at short phase heights demonstrates the decrease in sensitivity to longer wavelengths (where the planet is brightest). Finally, panel (vi) of Figure \ref{idealvsMKID} shows the affect of step 6. This step is performed before step 7 because those photons are ignored by the readout before it has the opportunity to evaluate their phase.

\subsection{Analysis}
\label{Analysis}

{\textsc VIP} \citep{gonzalez2017vip} provides tools to perform HCI analysis such as differential imaging, exoplanet discrimination and plot contrast-separation curves. These tools were adapted to MEDIS in order to implement the described post-processing techniques and data visualization.

For example, speckles with lifetimes larger than the frame integration time can be mitigated with differential imaging techniques whereby some fixed property of the system is exploited to generate and subtract a reference speckle field using principle component analysis or a similar algorithm \citep{lafreniere2007new, soummer2012detection}. ADI \citep{marois2006angular} and SDI \citep{sparks2002imaging} use a differential motion (azimuthal and radial, respectively) of the pupil compared to the sky.

The photon counting nature of MKIDs and the essentially zero read-noise enable photon-timing-statistics-based exoplanet discrimination methods, which are implemented in MEDIS. Dark Speckle Imaging (DSI, \cite{labeyrie_images_1995}) looks for regions of zero counts in the focal plane exploiting the fact that locations containing companions will be prohibited from reaching such levels of flux because of the companion's PSF. MEDIS has the functionality to run DSI and stochastic speckle discrimination alone or in combination with differential imaging techniques. 




The $5\sigma$ contrast-separation curve is a common metric of sensitivity in exoplanet direct imaging. Here we define contrast as 

\begin{equation}
C(r) = \frac{d}{I^{*}}\frac{k(r)\delta_\textrm{rms}(r)}{T(r)},
\label{eq:contrast}
\end{equation}
where $d$ is the detection threshold (usually 5), $\delta_\textrm{rms}$ is the azimuthal noise, $k$ is a penalty factor, $I^*$ is the brightness of the unocculted star, and $T$ is the algorithmic throughput. $T$ is an important measurement because different post-procesing algorithms will decrease the amplitude of companions at different locations. 
$\delta_\textrm{rms}(r)$ is calculated by taking the standard deviation of resolution elements of size $\lambda/D$ organized into rings around middle of the image. These data are then smoothed with a Savitzky-Golay filter. The limited sample statistics at small angles decreases the confidence level of detection (CL). Therefore to maintain an equal CL at all radii, a penalty factor of $k(r) = \sqrt{1 + 1/(n-1)}$ is applied, where $n$ is the number of resolution elements in each ring \cite{mawet2014fundamental}.

\section{MKID Parameter Investigation}
\label{sec:results}

\begin{figure*}
    \begin{minipage}[b]{\textwidth}
            \includegraphics[width=\columnwidth]{{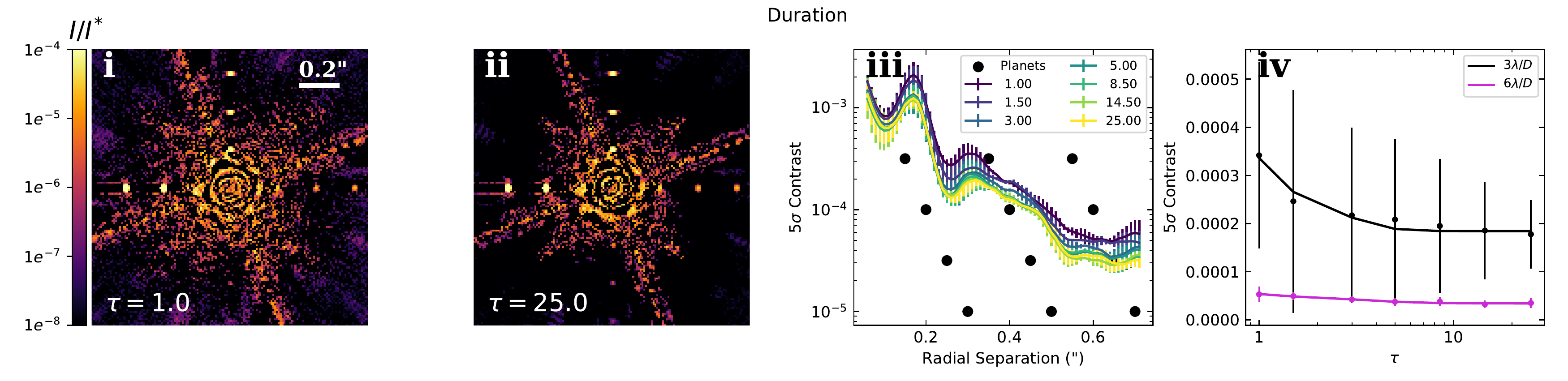}}
    \end{minipage}
    \begin{minipage}[b]{\textwidth}
            \includegraphics[width=\columnwidth]{{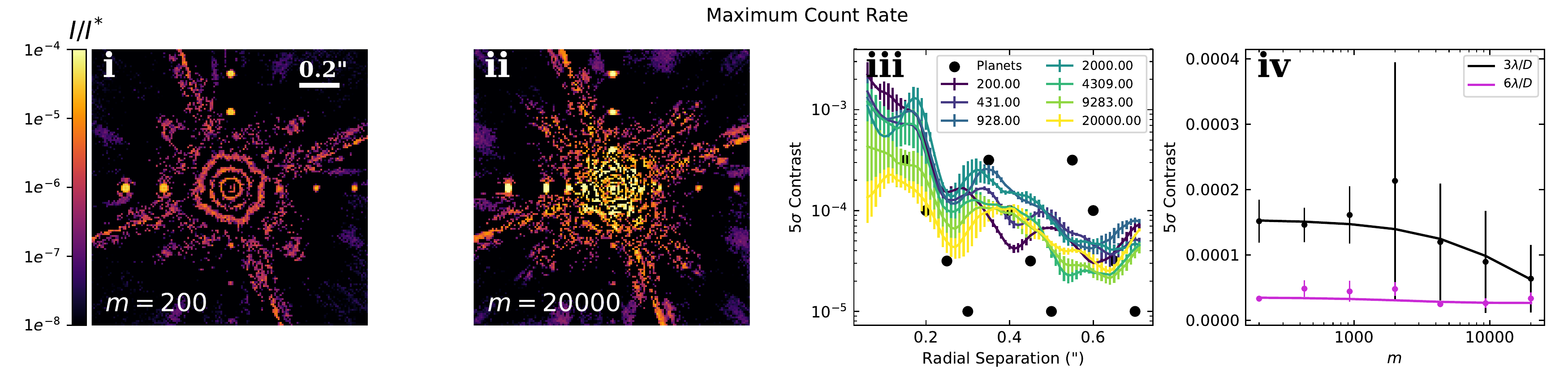}}
    \end{minipage}%
    \begin{minipage}[b]{\textwidth}
            \includegraphics[width=\columnwidth]{{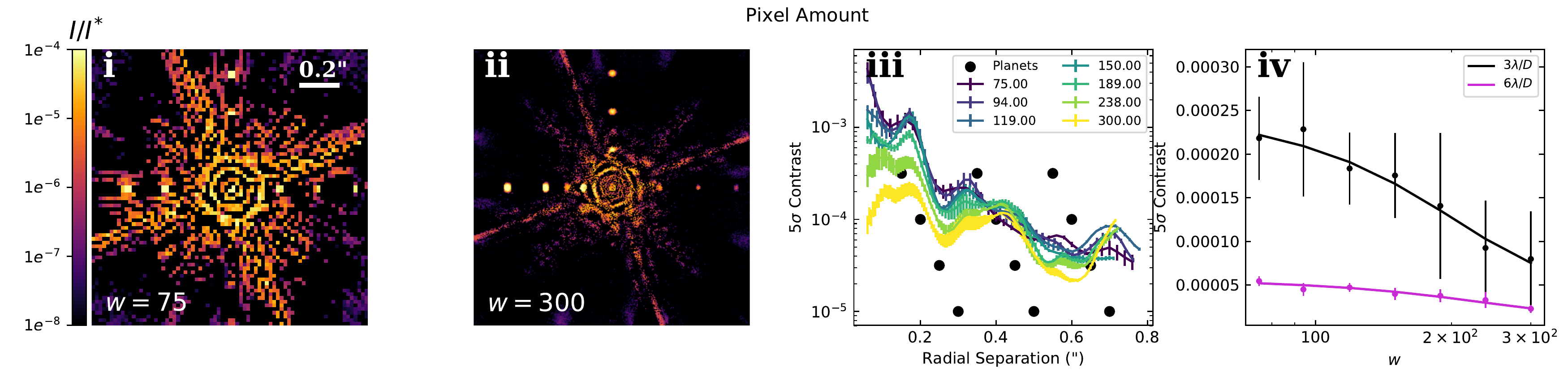}}
    \end{minipage}

    \caption{SDI analysis performed on MKID observations with different values for the observation duration (top), maximum count rate (middle), and array size (bottom). Columns (i) and (ii) show the reduced images at the lowest and highest parameter value under investigation. The companions are at contrast levels 10$^{-3.5}$, 10$^{-4}$, 10$^{-4.5}$ and 10$^{-5}$. The (iii) column is the 1D 5$\sigma$ contrast separation curve from observations with no companions and colored according to the value of the parameter under investigation. Column (iv) shows the relationship of contrast with the MKID parameter at two separations.}
    \label{array_size}
\end{figure*}
%
\begin{figure*}
 
    \begin{minipage}[b]{\textwidth}
            \includegraphics[width=\columnwidth]{{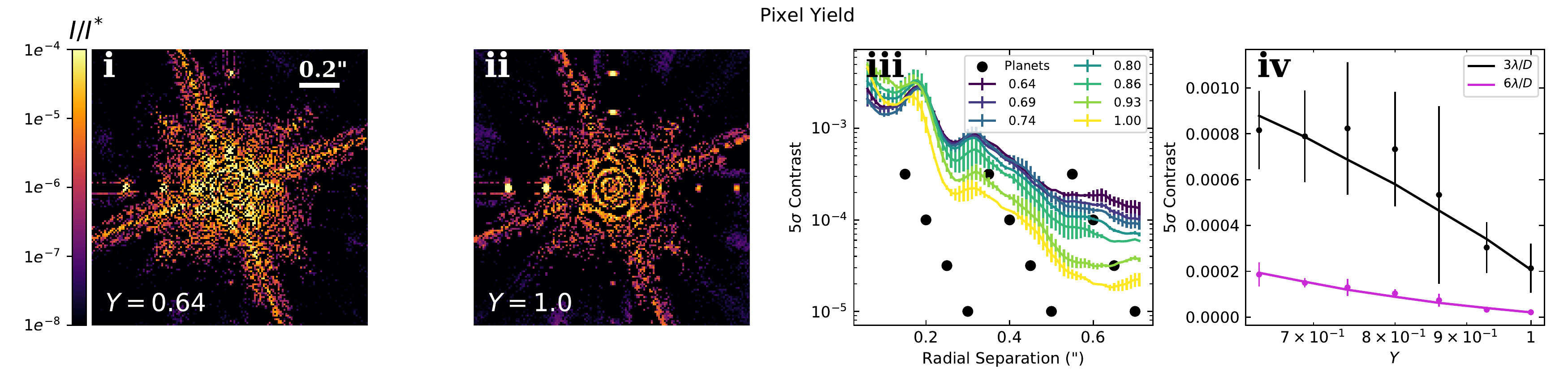}}
               \end{minipage}%
    \begin{minipage}[b]{\textwidth}
            \includegraphics[width=\columnwidth]{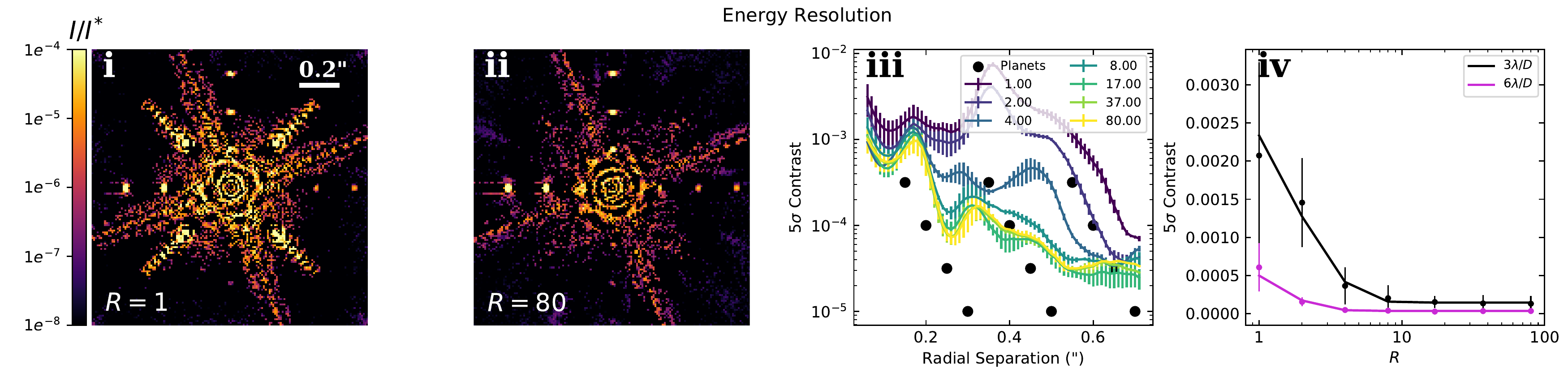}
    \end{minipage}%
    \begin{minipage}[b]{\textwidth}
           \includegraphics[width=\columnwidth]{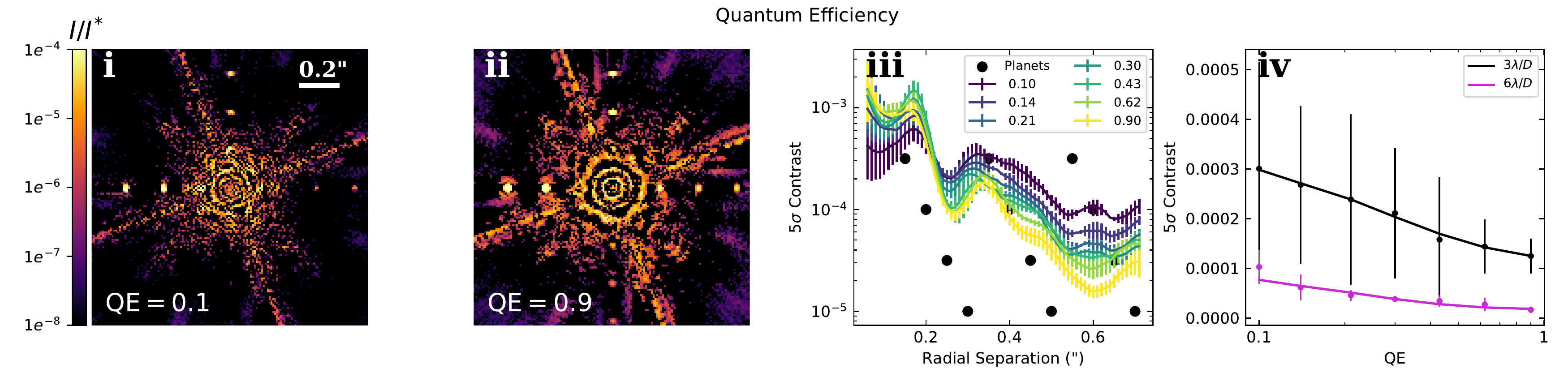}
    \end{minipage}%
    \caption{SDI analysis performed on MKID observations with different values for the pixel yield (top), mean energy resolution (middle), mean quantum efficiency (bottom). The figure has the same format as Figure \ref{array_size}} 
    \label{g_mean}
\end{figure*}

            


In order to demonstrate the capabilities of MEDIS we chose to investigate the effect of changing observation or MKID-specific parameters independently above and below a median value roughly equal to current devices. The parameters we investigated were the total duration of the observation $\tau$, maximum count rate $m$, pixel amount $w$, pixel yield $Y$, mean spectral resolution $R$, and mean quantum efficiency QE.

\subsection{Methods}

MEDIS was used to generate a series of simulated observations for a typical high contrast imaging system with an MKID camera of varying properties. The parameters used in these simulations are given in Tables \ref{tab:simulation} -- \ref{tab:mkids}, with the exception of some of the MKID parameters that varied between runs in order to investigate their effect. It is assumed there are no inter-dependencies between the parameters under investigation. 

There is no dithering and stacking of frames, which would give the mosaic a higher pixel yield, but may also decrease the mean $R$ if pixels with different values of $R$ are combined by degrading the spectra of higher $R$ pixels to their worse performing counterparts. That study is saved for later work.

Each observation was conducted and processed with and without companions. This allowed $\delta_\textrm{rms}$ to be measured accurately at all separations by removing the requirement to mask the companions or measure only part of the annuli (the difference in the noise between the observations with and without companions is negligible). 

$T$ was measured by comparing the flux before and after processing of twelve companions at a range of separations and magnitudes and a black body spectra of 1500K for an average age brown dwarfs. The companions are at contrast levels 10$^{-3.5}$, 10$^{-4}$, 10$^{-4.5}$ and 10$^{-5}$ at left of center, above center, right of center, below center, respectively. Since $T$ depends on the input flux of the companion it was measured for the four input fluxes at range of separations. The average of these four measures was then interpolated onto the sampling of the noise measurement. 

Each MKID parameter under investigation was tested three times with different realisations of atmosphere and aberration maps simulating variations between observing nights. All the observations were binned into a series of spectral datacubes, SDI analysis performed on each timestep and then median collapsed to yield 2D images.  


The (i) and (ii) columns of of Figures \ref{array_size} \& \ref{g_mean} show the results for the lowest and highest value of the test parameter, respectively. The contrast separation curves for all the nights are shown in the (iii) column along with markers for the locations of the planets. The (iv) column shows the trend of the contrast with the increasing test parameter taken at 3$\lambda/D$ and 6$\lambda/D$. These trends are fit with a exponential curve.

\subsection{Results and Discussion}

The top row of Figure \ref{array_size} shows the effect of observation duration on test data. When comparing the two images it is evident that the noise is slightly higher in the $\tau =1$ measurement (two time samples) than in the $\tau =25$ image (from fifty time samples), while the final brightness of the companions is relatively constant between images, indicating an increase in contrast performance for higher $\tau$. This trend is reflected in the (iii) panel by the higher $\tau$ curves having lower values of contrast relative to the black dots that indicate the values required to detect the planets. This increase in $\tau$ was sufficient to enable the detection of the planet at $10^{-4.5}$ contrast and 0.65" separation. The trend of contrast improvement with $\tau$ also appears to be very constant as indicated by the roughly monotonic decrease in contrast in both curves of panel (iv). The exponential decrease in contrast on this timescale is to be expected based on the speckle lifetime of uncorrected atmospheric speckles $\tau_{atmos}=0.6D/\bar{v} = 0.48s$, where $D$=8 is telescope aperture diameter and $\bar{v}$=10 is the mean wind speed \cite{macintosh2005speckle}. The effect is larger at closer separations -- the increase in $\tau$ for the range of values tested resulted in a contrast improvement of $\sim$2 at $3\lambda/D$ but $\sim$1.5 at $6\lambda/D$. Furthermore, this trend plateaus around $\sim$5 s at both separations, which is to be expected from the correlated nature of speckle noise. 


As seen in the second row of Figure \ref{array_size}, a low maximum count rate has severe consequences on the throughput of companion objects at close separations. This is to be expected as the PSF is brightest here and any information on the presence of companion objects is lost during the saturation of the affected pixels. Then, as parent star PSF is subtracted through differential imaging, there is no planet signal remaining. Conversely, a high count rate enables the detection of companions at these close separations. This trend is best exemplified in panel (iv) where the 3$\lambda/D$ curve is trending downwards whereas the 6$\lambda/D$ curve is comparatively flat. 


The bottom row of Figure \ref{array_size} paints a similar story. On first observation, increasing the pixel count while keeping the FoV of the device constant means a more finely sampled focal plane leading to a better PSF subtraction, but correspondingly decreases the throughput of the planets. This can be seen in the S/N of the $10^{-5}$ companions in the 75$\times$75 grid when compared to the 300$\times$300. Therefore the contrast stays roughly similar at wide separations. However, at close separations the final sampling meant less pixels saturated leading to a better PSF subtraction, enabling the detection of these challenging companions. 

In the top row of Figure \ref{g_mean} it is shown that increasing the pixel yield actually had a relatively large effect as both the noise and the throughput measurements benefit. It appears as though a sensor array with a larger fill factor leads to a more accurate PSF model and therefore less self subtraction of companion objects. For example, an increase in $Y$ from 64\% to 100\% produces an improvement in contrast performance of a factor of nearly 4 at $3\lambda/D$. There is also a clear monotonic performance improvement at 3$\lambda/D$. At the closest separation there is not much of a difference owing to the saturation of the pixels at all yields.




An increase in $R$ from what's achieved with the current MKID arrays does not improve the contrast when using SDI alone on this investigation. Each pixel is measuring the phase height accurately enough to place each photon in the correct bin of sixteen spectral samples. The bottleneck here is the chromaticity of the optics. There is, however, a very severe impact on contrast at values of $R$ lower than current capabilities. This is best demonstrated by looking at the smearing of the satelite speckles in the $R=1$ reduced image. 



The final row of Figure \ref{g_mean} shows the effect of QE on contrast. In panel (ii) it is clear that the benefit of increased sensitivity at smaller separations is wasted because of the saturating pixels. Further out though, the companions are appreciably brighter. Each of the $10^{-5}$ companions can be clearly identified by eye (the telescope spiders also increased in brightness which increased the $\delta_\textrm{rms}$, however the spiders could be safely masked to enhance the S/N of the companions). This highlights the importance of sensitivity in high contrast imaging of faint companions.






\section{Conclusion}
\label{sec:conclusions}

We developed a simulation package capable of providing pragmatic predictions from MKID-based high contrast instruments to demonstrate their potential. Using this simulator we showed that even when applying simple SDI on a standard HCI instrument, and not exploiting the photon counting capabilities of MKIDs, a contrast ratio of $10^{-4}$ and $10^{-4,5}$ at $3\lambda/D$ and 6$\lambda/D$ is attainable from a <30 s observation. We also showed that increasing either QE or pixel-yield can have a large impact on the identification of low flux companions at wider separations. In practice, the high-sensitivity nature of MKID instruments means that neutral-density filters are employed to avoid saturating the detectors, which decreases the amount of planet signal from pixels where saturation is not a concern. This suggests that some method of preventing the saturation of the central pixels while maintaining all of the flux further out will help detect fainter companions around bright stars. This could either be through better coronagraphy, checks in the readout firmware or software, or by increasing the maximum count rate or pixel sampling of the focal plane. This also provides further support that exploiting the information encoded in the photon arrival time statistics with techniques, like single-photon SSD, is essential for high-contrast instruments to capture low flux companions.

\section*{Acknowledgments}
Rupert Dodkins is supported by the National Science Foundation award number 1710385. Kristina K. Davis is supported by an National Science Foundation Astronomy and Astrophysics Postdoctoral Fellowship under award AST-1801983. Isabel Lipartito is sup-ported  by  the  National  Science  Foundation  GraduateResearch Fellowship under grant number 1650114


\section*{Appendix}
\label{sec:appendix}

What follows is a summary of some of the main configuration parameters that the user controls. 

\subsection{Simulation-wide parameters}
\label{sec:simulation}

\begin{table}
\caption{Examples of simulation-wide configuration parameters.}
\label{tab:simulation}	
	\centering
	\begin{tabular}{lc} 
		\hline
		Parameter & Example\\
		\hline
		\py{save_to_disk} & \py{True} \\
        \py{num_processes} & \py{8} \\
        \py{debug} & \py{False}\\
        \py{verbose} & \py{True}\\
        \py{memory_limit} & \py{10}\\
        \py{checkpointing} & \py{100}\\
        \py{save_photontable} & \py{True}\\
        \py{save_rebinnedcube} & \py{True}\\
		\hline
	\end{tabular}
\end{table}

Some of the main simulation-wide parameters are shown in Table \ref{tab:simulation}. Frequently a user will run several simulations in quick succession to observe differences in plots that appear at the end of the simulation. The \py{save_to_disk} parameter gives the user the option to skip the time overhead of saving to disk and clutter if it is not required. 

The \py{num_processes} parameter is the number of worker processes to use for any of the places where multiprocessing is used.The main time when this is used is in \py{Propagation} when the time steps are time independent (open-loop AO with no servo delay). 


In the \py{debug} and \py{verbose} modes additional plots and messages are generated as the simulation progresses. This includes the distributions used in generating the realisations of \py{Propagation} and \py{Quantization}. Debugging plots can also be individually toggled for specfic functions in the calls to those functions.

The \py{memory_limit} is the amount of dynamic memory in GB that can be used to store time samples simultaneously before chunking is employed. The save file is then populated on the fly rather than one event at the end of the simulation. Similarly, \py{checkpointing} tells the simulation at what time step to store the data on the fly. The minimum of \py{max_chunk} and \py{checkpointing} is used to determine when to store the time chunks.

The \py{save_photontable} and \py{save_rebinnedcube} parameters determine whether the quantized photon-lists are stored to disk in a format that can be read by the MKIDPipeline and the photon-lists are rebinned into a tensor and stored ready to be processed with conventional differential imaging techniques. 

The input/output parameters class controls the location of all the atmosphere and aberration maps, reference spectra, cached fields, the realisations of the MKID device properties, photons and reduced science data. The default is for all this data for a single investigation to be encapsulated in the test directory.


\subsection{Propagation parameters}
\label{sec:prop_params}

\begin{table}
\caption{Examples of Propagation configuration parameters.}
\label{tab:WavefrontProp}	
	\centering
	\begin{tabular}{lc} 
		\hline
		Parameter & Example\\
		\hline
		\py{duration} & \py{1} s\\
		\py{n_times_init} & \py{100}\\
		\py{n_times_final} & \py{200}\\
		\py{FoV} & \py{1.4} as\\
		\py{grid_width} & \py{512}\\
        \py{beam_ratio} & 0.25\\
		\py{num_objects} & \py{2}\\
		\py{wvl_range} & \py{[800, 1500]} nm\\
		\py{n_wvl_init} & \py{8}\\
		\py{n_wvl_final} & \py{16}\\
		\py{maskd_size} & \py{256}\\
		\py{focused_sys} & \py{True}\\
        \py{save_planes} & \py{[add_atmos, deformable_mirror]}\\
		\hline
	\end{tabular}
\end{table}



Some of the Propagation parameters are shown in Table \ref{tab:WavefrontProp}.  The \py{duration} and \py{num_frames} (number of frames) control the sampling in the time domain. The \py{FoV} (field of view), \py{grid_width}, and \py{beam_ratio} control the sampling in the spatial domain. If \py{maskd_size} < \py{grid_size} a crop with diameter \py{maskd_size} will be applied at the final focal plane. 

The \py{wvl_range} (wavelength range) and \py{n_wvl_init} control the sampling in the wavelength domain. The wavelength and time dimensions can later be interpolated to the final desired sampling to save time over the generation and propagation of new \py{wavefront}s by setting \py{n_wvl_final} > \py{n_wvl_init} and \py{n_times_final} > \py{n_times_init}.

The \py{num_objects} parameter controls the number of astronomical objects. The \py{save_planes} controls the location of surfaces in the optical train where the eletric fields are stored. It is a list of any function that manipulates a \py{wavefront}, or any labelled surface in the prescription, that when used will trigger MEDIS to populate \py{fields} with the electric field at that point. If \py{save_planes} is set to \py{None} then the field will only be saved during the final detector function at the end of the telescope chain.

The sampling in the focal plane must be the same for each wavelength. If the optical system is in focus then this is achieved by using the same beam ratio for each wavelength. On the other hand if \py{focused_sys} is \py{False} a linear scaling factor is applied to the \py{beam_ratio} to spatially descale the final image.

\begin{table}
\caption{Examples of telescope configuration parameters.}
\label{tab:telescope}	
	\centering

	\begin{tabular}{lc} 
		\hline
		Parameter & Example\\ 
		\hline
		\py{prescription} & \py{`general'}\\
         \py{entrance_d} & \py{8} m\\
         \py{use_atmos} & \py{True}\\
         \py{aber_params}  & 
         \begin{tabular}{@{}l@{}}
                 \py{`QuasiStatic': False,}\\
                 \py{`Phase': True,}\\
                 \py{`Amp': False,}\\
                 \end{tabular}\\
         \py{use_ao} & \py{True}\\
         \py{ao_act} & \py{64}\\
         \py{open_loop} & \py{True}\\
         \py{servo_params} & \py{[0.1, 0]} ms\\
         \py{use_astrogrid} & \py{True}\\
         \py{occulter}  & \py{`Vortex'}\\
         \py{rot_rate} & \py{0}\\
         \py{use_spiders} & \py{True}\\
		\hline
	\end{tabular}

\end{table}

The \py{use_astrogrid} parameter tells the DM to form two orthogonal sinusoids resulting in four satelite speckles, which is a useful technique in HCI for photometry and astrometry when a coronagraph is occulting the primary star. These speckles can be seen in the focal planes of Figure \ref{maps}.

\py{rot_rate} allows a simulation of an observation in pupil stablized mode, which is where the FoV derotator is turned-off. This is useful for applying techniques like Angular Differential Imaging.

\py{open_loop} determines if the AO is operated in open or closed loop mode. The two elements of \py{servo_params} correspond to servo lag and integrated frame period. The \py{Telescope} class will check if these parameters conflict with time parrallelisation request that requires the times to be independent.

For the \py{`Solid'} coronagraph, the occulter is a simple binary mask that only blocks light out to a certain radius. For \py{`Gaussian'}, the occulter has a graded transmission that follows a Gaussian profile. This means that the light is concentrated to the edges of the pupil much more effectively. The \py{`8th\_Order'} coronagraph can concentrate all of the parent star light the region blocked by the Lyot stop for 100\% rejection of the parent star light \citep{kuchner2005eighth}. The \py{`Vortex'} option is a vector vortex occulter for high throughput and small inner working angle coronagraphy. This coronagraph was implemented by adapting code from the HEEPS pipeline \citep{ChristianDelacroix}. 


Some of the main astrophysics parameters are shown in Table \ref{tab:astro}. If \py{star_spec} is an integer then the star electric fields are scaled relative to brightness of a blackbody at that temperature in Kelvin. Reference star emission spectra can also be loaded and used. For the investigation presented here, the black body assumption allows the spectrum to be sufficiently sampled with 8 points across the full band.

Together with the total intensity of \py{fields}, \py{star_flux} determines how many photons are created during the \py{Quantization} step across the whole array at that timestep (the process of generating photons is described in detail in Section \label{detector}). The default value corresponds to faint exoplanet target star or one of typical brightness but with some neutral density filtering applied to avoid saturating the detector. 


\begin{table}
\caption{Example astrophysics configuration parameters.}
\label{tab:astro}	
	\centering
	\begin{tabular}{lc} 
		\hline
		Parameter & Example\\ 
		\hline
        \py{star_spec} & \py{4000} K\\
        \py{star_flux} & \py{1e6} /s\\
        \py{use_companion} & \py{True}\\
        \py{contrast} & \py{[1e-3.5, 1e-4, 1e-4.5, ...]}\\
        \py{companion_xy} & \py{[[2,0],[0,2.5],[-3,0],...]}$\lambda/D$\\
		\hline
	\end{tabular}
\end{table}

The \py{contrast} and \py{companion_xy} parameters control the brightness and location of the companions respectively. The companion spectra is assumed to be constant over the J and H bands used in the investigation.



 


\begin{table}
\caption{Examples of the atmosphere configuration parameters. }
\label{tab:atmos}	
	\centering
	\begin{tabular}{lc}
		\hline
		Parameter & Example\\ 
		\hline
        \py{model} & \py{`single'}\\
        \py{r0} & \py{0.4}\\
        \py{cn} & \py{1*1e-14}\\
        \py{L0} & \py{10}\\
        \py{v} & \py{10}\\
        \py{h} & \py{10}\\
		\hline
	\end{tabular}
\end{table}


MEDIS uses HCIpy for atmospheric phase error map generation. The user chooses the number of layers, characteristic coherence lengths, altitudes ($h$), wind velocities and directions, as well as the refractive index structure parameters $C_n^2(h)$. that define the layers. The low frequency turbulence is responsible for the positioning of the centroid in the focal plane, which has a large impact on the performance of small inner working angle coronagraphs. However, modeling both the large and small scales simultaneously for each phase screen is very resource-intensive. HCIpy compensates for the low-frequencies in the Kolmogorov spectrum by incorporating sub-harmonics of the aperture to more finely sample the Fourier space close to the origin \citep{lane1992simulation}.

The different options for \py{model} are \py{`single'}, \py{`multiple'}, \py{`hcipy_standard'}. \py{`hcipy_standard'} ignores the remaining parameters in this class and instead uses the default ones provided by HCIpy.

\subsection{MKIDs parameters}

\begin{table}
\caption{Examples of the main MKID device configuration parameters.} 
\label{tab:mkids}	
	\centering

	\begin{tabular}{lc} 
		\hline
		Parameter & Example\\ 
		\hline
		\py{platescale} & \py{10} mas\\
        \py{device_shape} & \py{[150,150]}\\
        \py{dead_pix} & \py{True}\\
        \py{pix_yield} & \py{0.8} \\
        \py{hot_pix} & \py{True}\\
        \py{num_hot_pix} & \py{100}\\  
        \py{hot_bright} & \py{40e3}\\
        \py{hot_spec} & \py{`uniform'}\\
        \py{dark_frac} & \py{0.5}\\
        \py{dark_bright} & \py{40}\\
        \py{dark_spec} & \py{`Gaussian'}\\
        \py{phase_uncertainty} & \py{True}\\
        \py{phase_background} & \py{True}\\
        \py{R_mean} & \py{8}\\
        \py{R_sig} & \py{3}\\
        \py{QE_vary} & \py{True}\\
        \py{QE_mean} & \py{0.2}\\
        \py{QE_sig} & \py{0.04}\\
        \py{remove_close} & \py{True}\\
        \py{dead_time} & \py{2e-5} s\\
		\hline
	\end{tabular}
\end{table}

The \py{device_shape} parameter is inline with that of MEC, which has a larger than 20 kilopixel array. The number of functioning pixels is however lower than this value. Pixels are lost because of fabrication uncertainties leading to the spectral profile of two or more pixels overlapping, pixels having low responsivity, or pixels being driven with too much readout power \cite{dodkins2017bmkid}. The \py{dead_pix} Boolean dictates whether or not to apply the \py{pix_yield} parameter. 

The \py{hot_pix} and \py{dark_pix} Boolean's control whether to apply these non-astrophysical photons. The brightness and distributions of both types are controlled by the user. In Figure \ref{idealvsMKID}iv the count rates of hot pixels are arbitrarily set to be much larger than dark counts that they far out-number the dark counts even though the number of affected pixels are far fewer. However, hot pixels are relatively simple to identify and remove \cite{mchugh_readout_2012}.

The \py{phase_uncertainty} controls whether or not to apply the random bias (from the $R$ distribution) to each phase height measurement and the \py{phase_background} controls whether to compare the measured phase height to the pixel background or just allow all photons to be detected. The \py{R_mean} and \py{R_sig} control the distribution across the array at the lowest wavelength.

The \py{remove_close} and \py{dead_time} parameters can cause the simulation to take vastly longer times as the simulation iterates through each pixel's photon list to determine which ones should be registered and can be the longest part of the photon quantization process by {\it Quantizatiion}.  

\bibliographystyle{yahapj}
\bibliography{library}

\end{document}